\documentstyle[aps,prl,epsf,twocolumn]{revtex}
\begin{document}
\def\bea{\begin{eqnarray}}
\def\eea{\end{eqnarray}}
\def\d{\delta}
\def\nn{\nonumber}
\def\r{\rho}
\def\la{\langle}
\def\ra{\rangle}
\def\e{\epsilon}
\def\n{\eta}

{\bf {\noindent Comment on ``A simple one-dimensional model of heat
conduction which obeys Fourier's law''}}

In a recent letter Garrido et al~\cite{garr} consider heat conduction
in a model of hard point particles of alternating masses on a
one-dimensional line. Based on a number of numerical results, the authors
claim that this momentum conserving model exhibits Fourier's law. We first
comment on the apparent contradiction with an earlier
result of Prosen and Campbell~\cite{pros} (PC). We  then point out certain
inconsistencies in their results  and disagreements with our own results. 

The authors have first measured the system size dependence of the mean
current $ \la J \ra = \la N^{-1}\sum_l m_l u_l^3/2 \ra$ 
where $m_l$, $x_l$ and $u_l$ denote the mass,
position, velocity of the $l$th particle and $N$ is the number of
particles . As they correctly point  
out, it is not possible to  make definite conclusions from this
simulation data, since the asymptotic regime may not have been
attained.
Next the authors compute the  current-current correlation
$C(t)= N \la J(t) J(0) \ra  $  
and find a decay $ C(t)\sim t^{-1.3}$, 
which is sufficiently fast to give a finite Kubo conductivity $\kappa$. 
This would seem to contradict the exact result of PC
on infinite thermal conductivity in momentum conserving 
systems. Their proof applies to this model. 
This apparent contradiction can be explained by the fact that, in their
simulation, Garrido et al work in an ensemble with 
total momentum set to zero and in this case the PC 
proof does not predict anything. 
As has been pointed out by Bonetto et al~\cite{bone}, the correct expression for
the Kubo formula requires one to use the connected part of the
correlation function if one is working in the canonical
ensemble~\cite{lenn}. Alternatively one may fix the momentum to be zero and work
with the usual correlation function as Garrido et al have done. 
Thus the work of PC does not really prove divergence of $\kappa$. 

However some  aspects of the paper appear to be  unsatisfactory and
need some explanation.  
Firstly the linear temperature profiles
obtained in the paper are inconsistent with the finding of finite
$\kappa$. The temperature dependence of
$\kappa$ can be scaled out from the Kubo formula, yielding a
$T^{1/2}$ dependence (given $\kappa$ is finite). This
follows from the fact that the correlation $C_T(t)$ at
temperature $T$ has the scaling form $C_T(t)=T^3 C_1(T^{1/2}t)$. The
$T^{1/2}$ dependence of $\kappa$ also follows from simple kinetic theory
arguments. A temperature dependent $\kappa$ at once leads to nonlinear
temperature profiles.  Infact in our study of the same model 
~\cite{dhar} we  clearly see  the expected nonlinear profiles for
similar system sizes. 
One possible reason for the difference could be that Garrido et al use determistic
heat baths unlike the stochastic heat baths used by us. It is not
clear how well such deterministic baths simulate true thermal
sources. Another possibile  source of error is the way Garrido et al
define local temperature, namely, by  measuring the mean kinetic
energy and mean position of each particle. In
one dimensions position fluctuations are large ($\sim \sqrt{N}$) and  
a more correct procedure is the one used by us: to define the local  number and energy densities 
as  $n(x,t) = \la \sum_l \d(x-x_l) \ra $ and $\e(x,t) =
\la \sum_l (m_l u_l^2/2) \d(x-x_l) \ra $  respectively, 
and then define the local temperature as $T(x)= 2 \e(x)/n(x)$.

Secondly, from our own simulations, we are unable to verify the results
of~\cite{garr}. The authors have computed $C(t)$ and also the
onsite correlator $c(t)=\la j_i(t) j_i(0)  \ra$. They
find that for the unequal mass case, $C(t)$ and $c(t)$ have roughly the
same long-time decay $\sim 1/t^{1.3}$  while for the  equal mass case $c(t)
\sim 1/t^3$.  
Our results are summarized in Fig.~(\ref{cdec}) and the important
differences with ~\cite{garr} are: 

(i) We do not find any evidence for the  decay  $C(t) \sim 1/t^{1.3}$.
Infact the behaviour we find is consistent with the decay $J \sim
1/N^{0.83}$ found in~\cite{dhar}. However we do not wish to make a
strong case for the number $0.83$ though it is rather striking that we
get the same number from two very different approaches. 

(ii) The behaviour of $c(t)$ seems to be very different from that of
$C(t)$ contrary to what is claimed in~\cite{garr}. 
The authors comment that $c(t)$ has better averaging
properties and, because it shows approximately the same decay, this 
confirms the behaviour seen for $C(t)$. But is there any reason to expect
that for the unequal mass case $C(t)$ and $c(t)$ will have similar
decay laws ?    
Infact for the equal mass case we know that $C(t)$ is 
a constant (since $J$ is a constant of motion) and therefore the
behaviour of $C(t)$ and $c(t)$ are drastically different.

(iii) The equal mass
case is nonergodic since there are a macroscopic number of
conservation laws. Thus time averages from simulations are
initial-condition-dependent. We can verify this in our simulations
and also find that making the masses slightly unequal restores ergodicity. 
Thus  it is hard to understand how the decay $c(t) \sim 1/t^3$ is obtained by~\cite{garr}.     
We note that the paper by Jepsen~\cite{jep} (quoted by Garrido at al) only
gives $\la v_i(t) v_i(0) \ra \sim 1/t^3$. It is not clear how this
leads to the same prediction for the asymptotic behaviour of $c(t)$
which is more like $\la v_i^3(t) v_i^3(0) \ra $. Besides, the
calulation of Jepsen  is not done in the zero-momentum ensemble.   

In our simulations time averages were performed over $10^9-10^{10}$
collisions. As checks on our
simulations we found that $C(0)$ and $c(0)$ agree with their exactly known
values and also that $C(t)$ and $c(t)$ satisfy the exact scaling
relations mentioned above. In the equal mass case we got a constant
$C(t)$, as expected. 

In conclusion we have shown that there does not seem to be any hard
evidence to prove validity of Fourier's law in this system. 
We think it is as hard to conclude  
anything (about finiteness of $\kappa$ ) from correlation
function data as it is from the $J $versus $N$ data.

I thank Onuttom Narayan for illuminating discussions and for
critically reading the manuscript.

\noindent Abhishek Dhar\\
Raman Research Institute\\
Bangalore 560080\\
India.

\vbox{
\epsfxsize=8.0cm
\epsfysize=6.0cm
\epsffile{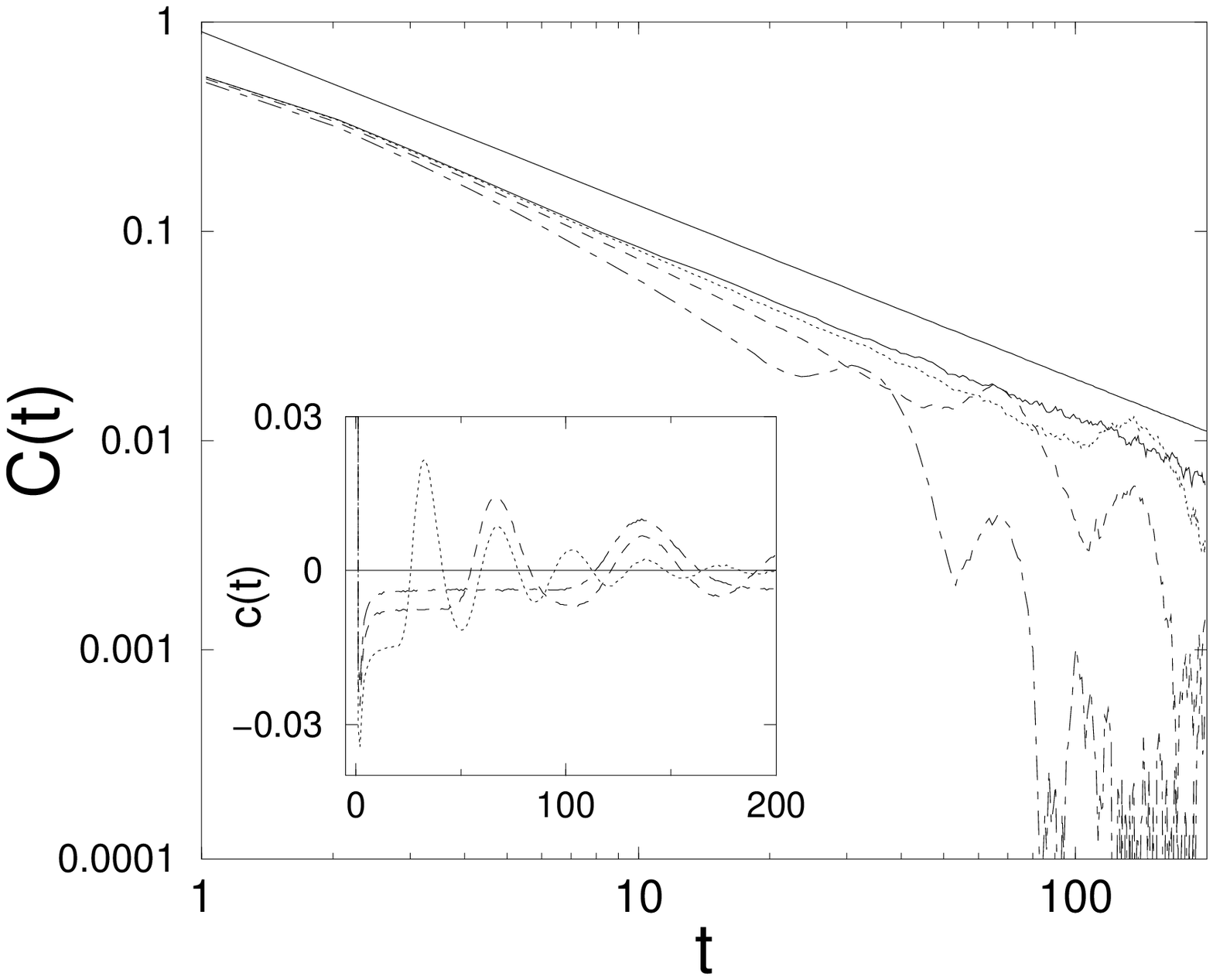}
\begin{figure}
\caption{\label{cdec} Plot of $C(t)$ for system sizes
$N=100$ (dot-dashed), $200  $ , $400 $ and $800$(solid) [ $T=1$, $m_1=1$, $m_2=2$]. Inset shows
$c(t)$ for $N=100$ (dots), $200$   and $400$ (dot-dashed) The
straight line corresponds to the power-law decay $\sim 1/t^{0.83}$.  
}
\end{figure}}


\begin{references}
\vspace{-1.5cm}
\bibitem{garr} P. L. Garrido, P. I. Hurtado and B. Nadrowski,
Phys. Rev. Lett. {\bf 86}, 5486 (2001). 
\bibitem{pros} T. Prosen and D. K. Campbell, Phys. Rev. Lett. {\bf
84}, 2857 (2000). 
\bibitem{bone}  F. Bonetto, J. L. Lebowitz and L. Rey-Bellet,
math-ph/0002052.  
\bibitem{lenn} For a derivation see ``Introduction to nonequilibrium statistical mechanics,
J.~A.~McLennan~(Prentice Hall, 1989), URL:http://www.lehigh.edu/~ljm3/ljm3.html''.      
\bibitem{dhar} A. Dhar, Phys. Rev. Lett. {\bf 86}, 3554 (2001).
\bibitem{jep} D. W. Jepsen, J. Math. Phys {\bf 6}, 405 (1965).
\end{references}
\end{document}